\documentclass{iopart}

\usepackage{graphics}

\begin{document}

\title{Rare $\beta$p decays in light nuclei}

\author{MJG Borge$^1$, LM Fraile$^2$, HOU Fynbo$^3$, B Jonson$^4$,
  OS Kirsebom$^3$, T Nilsson$^4$, G Nyman$^4$, G Possnert$^5$,
  K Riisager$^3$ and O Tengblad$^1$}

\address{$^1$ Instituto de Estructura de la Materia, CSIC, Serrano
  113 bis, E-28006 Madrid, Spain}
\address{$^2$ Grupo de F\'{\i}sica Nuclear, Universidad Complutense, 
  CEI Moncloa, E-28040 Madrid, Spain}
\address{$^3$ Department of Physics and Astronomy, Aarhus University,
 DK-8000 Aarhus C, Denmark}
\ead{kvr@phys.au.dk}
\address{$^4$ Fundamental Fysik, Chalmers Tekniska H\"{o}gskola,
  S-41296 G\"{o}teborg, Sweden}
\address{$^5$ Department of Physics and Astronomy, Ion Physics, Uppsala University,
  S-75120, Sweden}

\begin{abstract}
Beta-delayed proton emission may occur at very low rates in the
decays of the light nuclei $^{11}$Be and $^8$B. This paper explores
the potential physical significance of such decays, estimates their
rates and reports on first attempts to detect them: an experiment at
ISOLDE/CERN gives a branching ratio for $^{11}$Be of $(2.5 \pm 2.5)
\cdot 10^{-6}$ and an experiment at JYFL a 95\% confidence upper limit
of $2.6 \cdot 10^{-5}$ for $^8$B.
\end{abstract}


\pacs{23.40.Hc, 27.20.+n, 21.10.Gv}

\submitto{\jpg}

\section{Introduction}
The lightest chemical element in which beta-delayed proton ($\beta$p)
emission has been observed so far is carbon, in the decay of
$^9$C. This paper explores the possibility of seeing the decay mode
also in beryllium and boron, more specifically in the decays of
$^{11}$Be and $^8$B. These nuclei are the lightest (not counting the
deuteron) one-neutron and one-proton halo nuclei, respectively, and
the possible $\beta$p decays are intimately connected to this
structure. There are two reasons for this connection, the first being
the factorization of the wavefunction into a halo and a core part that
suggests ``independent decays'' of the two parts \cite{Nil00} which
naturally leads to final states with a continuum proton; we shall
discuss this model in the next section. The second connection is
specific for $\beta^-$ decays where one can derive the general
expression \cite{Jon01}
\begin{equation}
 Q_{\beta p}(^AZ) = (m_n-m_H)c^2 - S_n(^AZ) =\mathrm{782\, keV} - S_n(^AZ) 
\end{equation} 
from which it is seen that the $\beta$p decay only occurs for
nuclei with very low neutron separation energy. Similar relations
exist for beta-delayed deuteron and triton emission that are known to
take place close to the neutron dripline \cite{Nil00,Pfu11}. So
far $\beta$p decays have only been observed following $\beta^+$ or EC
decays. General recent reviews of these processes can be found in
\cite{Pfu11,Bla08}.

The following section discusses the characteristics of the proposed
decays in more detail including the estimated order of magnitude of
the decay rate. The expected branching ratios are quite low and some
general considerations are made on how they could be measured.
Sections 3 and 4 give details on first measurements of
the decays in $^{11}$Be and $^8$B from which upper limits on
the branching ratios can be given. The final section gives our
conclusions and the prospects of identifying the two decays.

\section{Expected rates}
A crucial feature of halo states is their intrinsic clustering so that
(a major part of) the wavefunction will factorize into a halo part and
a core part \cite{Rii00,Jen04}. Formally, the beta-decay can then be
thought of in terms of separate decays of the core (c) and halo (h)
parts \cite{Nil00}
\begin{equation}  \label{eq:fac}
  \mathcal{O}_{\beta} (|c\rangle |h\rangle) = (\mathcal{O}_{\beta}
  |c\rangle) |h\rangle + |c\rangle (\mathcal{O}_{\beta} |h\rangle) 
\end{equation}
but the actual final states will of course in general be
superpositions of the two terms on the right side e.g.\ to ensure they
have proper isospin. If the states on the right hand side are
close to being eigenstates for the final system one would expect a
large resemblance between the decays of the core nucleus and the halo
nucleus. Experimental confirmation of this idea may be found in the
decays of $^{12}$Be and $^{14}$Be, which are quite analogous
\cite{Jon04}.

\subsection{$^{11}$Be($\beta$p)}
The most recent mass tables \cite{mas11} give $Q_{\beta}= 11509.2 \pm
0.5$ keV for the decay of $^{11}$Be, which leaves 281 keV as the
energy window for beta-delayed proton emission.  Different assumptions
on the decay mechanism can give widely different rates \cite{Ang98}.
We shall therefore present several simplified models that lead to
rather straightforward expressions for the decay rate, and start by
considering sequential decays as described in R-matrix theory that can
handle broad intermediate states. However, the classical R-matrix
expressions may not give a sufficiently accurate description of the
decay mechanism. At the other extreme are models that consider decays
to proceed directly to continuum states, as has been assumed for most
calculations of the related $\beta$d decays \cite{Nil00,Pfu11}.

\begin{figure}
 \hspace{2cm}
 \resizebox{0.7\textwidth}{!}{\includegraphics{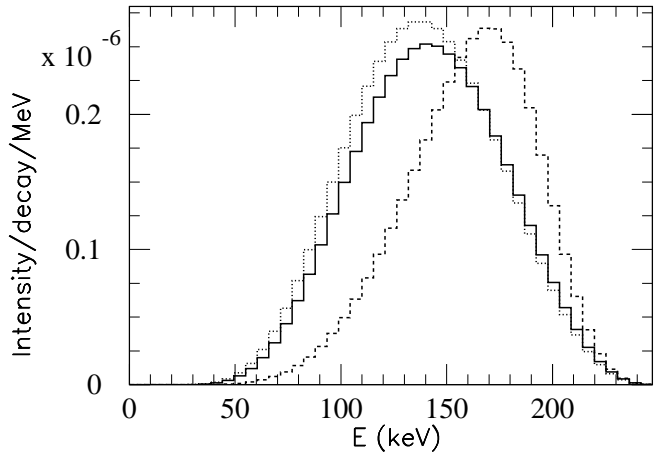}}
\caption{Energy spectra of beta-delayed protons from the decay of
   $^{11}$Be predicted in three different models presented in the text.
   Sequential decay through the 11.444 MeV state in $^{11}$B
   (dashed line), emission before decay (doted line),
   decay to continuum (full line). \label{fig:e11be}}
\end{figure}

Allowed beta decay of $^{11}$Be will populate $1/2^+$ and
$3/2^+$ levels in $^{11}$B, but since these would decay to $^{10}$Be+p
with s-wave and d-wave protons, respectively, only the $1/2^+$ states
can be expected to give a sizeable contribution. Out of the known
states \cite{Ajz90,Yam11} in the excitation energy range 10.0--12.5
MeV the only one that may have spin and parity $1/2^+$ is a state at
11.444 MeV with width 103 keV that has been seen to decay by emission
of $\alpha$ particles, but where the spin is so far undetermined.
The R-matrix expressions for $\beta$-decay have been derived for
several cases, see \cite{Bar89} and references therein, but we shall
here only consider a crude approximation based on the single-channel
single-level case in which the energy spectrum is:
\begin{equation}
  \frac{w_p}{w_{tot}} = T_{1/2} \frac{g_A^2}{K} f_{\beta}(Q-E)
  B_{GT}  \frac{\Gamma/(2\pi)}{(E-E_0)^2 + \Gamma^2/4}
  \frac{P(E-E_t)}{P(E_0-E_t)} \;, 
\end{equation}
where $w$ denote the decay rates for the proton branch and the total
beta decay, $T_{1/2}$ the experimental halflife, $g_A$ and $K$ the
standard weak decay constants \cite{Pfu11,Bla08}, $f_{\beta}$ the beta
decay phase space factor, $B_{GT}$ the reduced Gamow-Teller matrix
element squared, $E_0$ and $\Gamma$ the level energy and width, and $P(E-E_t)$
the proton penetrability ($E_t$ being the proton threshold
energy). Several energy dependent terms in the denominator are here
approximated as constants. To take into account that the level can
decay both by emission of $\alpha$ particles and protons one should
multiply by $\Gamma_p/\Gamma_{tot}$ to get the intensity of the proton
spectrum. This ratio is unknown, but may in the best case (if the
proton width is close to a single-particle width) reach $1/2$. The GT
strength $B_{GT}$ can at most be 3 (the value for a pure neutron to
proton decay). Inserting these maximum values the total branching
ratio becomes $1.0 \cdot 10^{-6}$, but a realistic value could be
lower by an order of magnitude or more. The predicted spectrum is
given in figure \ref{fig:e11be} for the case where the intensity (the
product $B_{GT}\Gamma_p/\Gamma_{tot}$) is a factor 50 below the
maximum, i.e.\ where the branching ratio is $2.0 \cdot 10^{-8}$.

As an intermediate step we consider next the suggested modification of
the R-matrix formalism containing the opposite time ordering
\cite{Bor93}, i.e.\ where the neutron is emitted before beta decaying
into a proton. As argued by Barker \cite{Bar94}, who presents a more
detailed derivation analogous to the way photon emission is treated,
this may be a simple way to effectively include the decays to
continuum states into the R-matrix framework.  In the limit where only
the ``emission before decay'' process is included the decay rate
becomes
\begin{equation}
  \frac{w_p}{w_{tot}} = T_{1/2} \frac{g_A^2}{K} f_{\beta}(Q-E)
  B_{GT}  \frac{1}{\pi} \frac{P(E-E_t)\gamma^2}{(S_n+E)^2} \;, 
\end{equation}
where $S_n$ is the neutron separation energy in $^{11}$Be and the
parameter $\gamma^2$ quantifies the neutron emission probability. For
a decaying level it is related to the level width through $\Gamma = 2
P \gamma^2$ and it must be less than the single particle width (the
Wigner limit, here 5.8 MeV). The total branching ratio is proportional
to $\gamma^2$ and is at most $2.5 \cdot 10^{-8}$. The predicted energy
distribution is also plotted in figure \ref{fig:e11be}.

Finally, the simple model for decays directly to continuum states
\cite{Rii90} gives
\begin{equation}
  \frac{w_p}{w_{tot}} = T_{1/2} \frac{g_A^2}{K} f_{\beta}(Q-E)
  B_{GT}(E) \frac{P(E-E_t)}{R} \frac{mc^2}{2\pi^2 (\hbar c)^2} \;, 
\end{equation}
where $m$ is the reduced mass of the outgoing proton and $R$ is the
channel radius used to evaluate the penetrability. The Gamow-Teller
strength is now explicitly energy dependent since the final state
wavefunction is a continuum state. The expression above results if one
uses a plane wave approximation for the final state when calculating
$B_{GT}(E)$.  If one instead uses a Coulomb wave the effect of the
transmission through the Coulomb potential is included in $B_{GT}(E)$
and one must replace the ratio $P/R$ by the
wavenumber $k$. The total branching ratio is in this latter case $2.3
\cdot 10^{-8}$, the corresponding energy distribution is given in
figure \ref{fig:e11be}.  More realistic calculations will use
distorted waves so that the final state strong interactions are also
included. A recent more sophisticated calculation \cite{Bay11} in a
two-body potential model gives a broad energy spectrum peaking at
0.1--0.2 MeV and a branching ratio of $3.0 \cdot 10^{-8}$. (This
specific calculation includes both Fermi and Gamow-Teller
contributions to the decay rate.)

To summarize, the different models predict somewhat different shapes of
the energy distribution and give a branching ratio that typically is a
few times $10^{-8}$, but could range up to $10^{-6}$. To see the decay
mode an experiment therefore needs to be sensitive down to the
$10^{-8}$ level, while considerably more intensity will be needed to
separate the different models.

\subsection{$^8$B($\beta$p)}
Turning now to $^8$B two facts make the decay directly to continuum
states less likely here: the smaller spatial extent of the halo in this
nucleus and the fact that there is an excited state in $^8$Be within
the $\beta$-decay window (actually the EC-decay window)
that the decay may pass through, namely the
$1^+$ $T=1$ state at 17.640 keV. It is situated 385 keV above the
proton threshold and is know to decay mainly by proton emission \cite{Til04}. If
equation (\ref{eq:fac}) is appropriate, we can estimate the matrix element of
the transition to be the same as for the ground state decay of $^7$Be
into the ground state of $^7$Li (the larger amplitude of the
wavefunction in $^8$B increases the rate by a factor 2.1 \cite{Bam77}
but this is compensated by a factor 1/2 from the isospin
Clebsch-Gordan coefficient squared). Both transitions are electron
capture decays for which the phase space goes as the Q-value squared,
so by scaling the halflife of $^7$Be we obtain an estimate of the
branching ratio of the $\beta$p transition of $2.3 \cdot 10^{-8}$.

Direct support for this estimate can be found from the three-cluster
calculations in \cite{Gri00} that predict a $B_{GT}$ to the 17.64 MeV
state of 1.366 and 1.997 for two different potentials employed. This
is in the same range as the $B_{GT}$ of 1.83 for the $^7$Be ground
state transition.

\subsection{Experimental considerations}
Direct detection of these $\beta$p decays is challenging since one has
to look for particles of low energy, with a low branching ratio, and
where the decay branch has to be identified e.g.\ through particle
identification (PID) techniques. As an alternative one may try to
detect the presence of the daughter nucleus or, more ambitiously, its
growing-in. In the cases in question here the final nucleus is stable
or longlived so one cannot make use of its decay, but must resort to
other methods. One possibility is accelerator mass spectroscopy (AMS)
as explained in the next section.

A low branching ratio by itself does not prohibit direct detection.
As an example, branching ratios well below $10^{-10}$ have been
detected for cluster radioactivity \cite{Pri89}. The challenge is
rather to identify the decay branch uniquely. For proton kinetic
energies below 1 MeV one could employ gas telescopes and use the
energy loss $\Delta E$ versus full energy $E$ to identify the
particle. However, the energy loss for a proton peaks just below 100
keV energy so the $\Delta E$-$E$ curves for protons, deuterons and
tritons will cross in this region, making separation e.g.\ between
protons and tritons (relevant for $^{11}$Be) impossible.
Identification via detection of $E$ and time-of-flight is an
alternative, but will typically have much less efficiency. Separating
protons from the background of energy loss signals from beta particles
could also be a problem. 

If the proton can be recorded in coincidence with the recoiling
nucleus this may be used to discriminate against other channels. This
``ratio cut'' method is useful to separate true events from response
tails of events at higher energy, as done in the $\beta\alpha$ decay
of $^{16}$N \cite{Azu94} and needed here for the case of $^8$B. It can
also be used to single out one decay channel among many others, as
done for the $^{11}$Li beta-decay to t+$^8$Li \cite{Mad09} and needed
here for $^{11}$Be. The main challenge with this method is to obtain
efficient detection of the recoil with sufficient accuracy in
energy, in the examples given the energy was higher than for our
$\beta$p decays.

\section{The $^{11}$Be experiment}
Four beta-delayed particle branches are energetically open for
$^{11}$Be, their Q-values are listed in table \ref{tab:QS}. We may
neglect the $\beta$n branch since it would proceed to the $^{10}$B
$3^+$ ground state and require a d-wave neutron with very low
energy. The $\beta$t branch leads to $^8$Be and would therefore give a
three-body final state, this could also be a quite interesting decay
mode. The main background for the $\beta$p branch will be
$\beta\alpha$ whose branching ratio was measured to be $3.1\pm0.4$\%
\cite{Mil82} (a recent determination gave $3.47\pm0.12$\%
\cite{Bus10}).

\begin{table}
 \caption{Q-values and separation energies in $^{11}$Be
 beta-delayed particle decays. \label{tab:QS}}
 \begin{tabular}{lcccc}
 \br
 $x$ & p & n & $\alpha$ & t \\ \mr
 S$_x$($^{11}$B) (keV)$^a$ & $11228.5 \pm 0.4$ & $11454.12 \pm 0.16$ &
 $8664.1 \pm 0.4$ & $11223.5 \pm 0.4$  \\
 $Q_{\beta x}$ (keV)$^a$ & $280.7 \pm 0.3$ & $55.1 \pm 0.5$ &
 $2845.2 \pm 0.2$ & $285.7 \pm 0.2$ \\
 \br
 \end{tabular}
$^a$Mass values from ref. \cite{mas11}
\end{table}


\subsection{The set-up and source production}
Due to the challenges involved in direct detection of the $\beta$p
decays we decided to explore the possibility of indirect
detection. The $\beta$p-daughter nucleus, $^{10}$Be, is radioactive
with a halflife of $1.39 \cdot 10^6$ years and is only present on earth
in minute quantities due to production by cosmic rays. 
It is used as a tracer in earth sciences
and procedures have been developed for detecting very small samples of
$^{10}$Be via accelerator mass spectroscopy (AMS), see e.g.\ \cite{Elm87,Mul08}.
Stated briefly, our experiment then consists in collecting a large
number of $^{11}$Be atoms, measure their number by detecting
$\gamma$-radiation from the $^{11}$Be decay and later measure the
number of produced $^{10}$Be atoms via AMS.

The $^{11}$Be source was collected in September 2001 at the ISOLDE
facility at CERN. The $^{11}$Be atoms were produced
by proton bombardement in a Ta target, ionized by laser ionization
\cite{Fed03}, accelerated and transported to the measuring station
where the ion beam passed two sets of collimators before being
implanted into a Be foil (Goodfellow, LS226536 L C, thickness 0.01mm,
purity 99.8+\%). 
The average intensity of the collected beam was $3.6\cdot 10^6$ 
ions/s, the collection took place for close to 40 hours.
The distance from the first collimator to the collection point was 177
mm. 
The transmission was optimized with a stable $^{23}$Na beam and a
radioactive $^{27}$Mg beam, the fraction of the activity that
deposited on the collimator was at most 12\%.

\subsection{Source strength}
Most of the decays of $^{11}$Be proceed through states in $^{11}$B that
decay by gamma emission, the most prominent $\gamma$-ray being the
2124.47 keV line with a total branching ratio, including feeding from
higher levels, of $b_{\gamma} = 0.355 \pm 0.018$ \cite{Mil82}.
The gamma detection was done with a Ge-detector
that was placed about 40 cm
from the collection point in the opposite direction of
the collimator. It was further shielded by 3 mm Pb to reduce the total
count rate. Still, the multi channel analyzer used for data taking had a
deadtime of $14\pm 1$\% determined from the ratio of live time and 
real time of the data taking system.

\begin{figure}
 \hspace{2cm}
 \resizebox{0.8\textwidth}{!}{\includegraphics{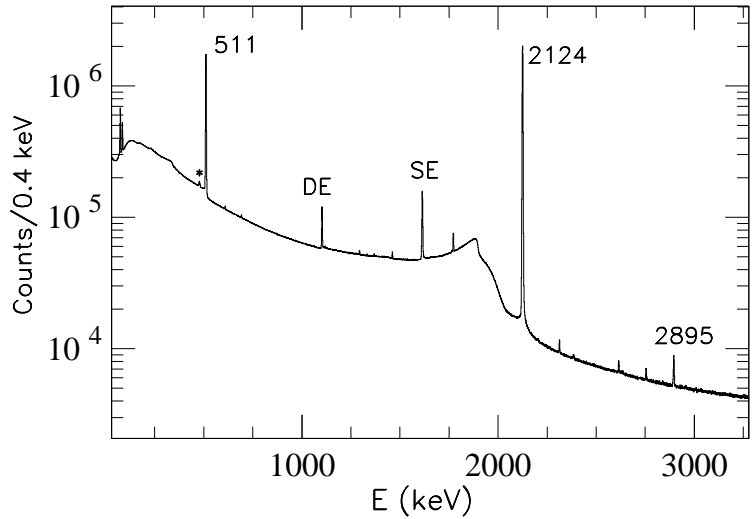}}
\caption{Gamma energy spectra recorded from the decay of
   $^{11}$Be, the number of counts per 0.4 keV is shown versus gamma-ray 
   energy. The star marks the 478 keV line. \label{fig:g11be}}
\end{figure}

The Ge detector was energy and efficiency calibrated \textit{in situ} with
absolutely calibrated sources of $^{60}$Co, $^{137}$Cs and $^{228}$Th
that bracket the energies from 239 keV to 2615 keV.
The total gamma spectrum from $^{11}$Be is shown in figure \ref{fig:g11be}.
The spectrum is dominated by gamma lines from the decay of $^{11}$Be
with a small background mainly from $^{40}$K and the $^{222}$Rn decay
chain. 
The 2124 keV line is clearly separated from background and gives,
after correction for detector efficiency and deadtime, a deduced amount
of $^{11}$Be atoms collected of $(5.2 \pm 0.3) \cdot 10^{11}$. As a
cross-check of this number we shall look at two other $\gamma$-lines
recorded from the $^{11}$Be decay. First the 2895 keV line that again
is  nicely separated in our spectra; we deduce an intensity
ratio with respect to the 2124 keV line of $0.241 \pm 0.011$\% in
agreement with the literature value of $0.227 \pm 0.008$\% \cite{Mil82}.
Secondly, we consider the 478 keV line that follows $\beta\alpha$
decays of $^{11}$Be to the first excited state in $^7$Li. The line is
therefore recoil broadened and care must be taken when extracting its
intensity since it furthermore is situated on top of a large background. Our
deduced intensity relative to the 2124 keV transition is $0.75 \pm 0.06$\%.

The literature value for this last ratio can be derived from the decay
scheme in \cite{Alb81} to be 1.13\% (it was not observed directly
since gamma detection was done with a NaI detector). However, as we
shall argue now, this value is most likely wrong.
Alternative determinations of this ratio
measured with Ge detectors can be extracted from experiments on the
decay of $^{11}$Li in which $^{11}$Be is fed. From two such recent
experiments accurate results can be obtained (although not published
previously): an experiment we did at ISOLDE \cite{Fyn04} gave
$0.67 \pm 0.11$\% and an experiment at TRIUMF \cite{Mat09,Sar09} gave
$0.69 \pm 0.07$\%. An unpublished recent $\beta\alpha$ experiment \cite{Bus10}
has revised the branching ratio of alpha particles to the 478 keV
excited state in $^7$Li to be $8.0 \pm 0.4$\% of all alpha decays, a result
based on a detailed analysis of the alpha spectrum but in disagreement
with the literature value of $12.6 \pm 1.2$\% \cite{Alb81}. Using the new
value one derives a 478/2125 ratio of $0.78 \pm 0.06$\%. Since
the three new determinations all are consistent with our
value for the ratio, we conclude that our spectra are internally
consistent which adds confidence to our derived source intensity.

\subsection{Source purity}
The collection foil must of course be free of previous $^{10}$Be
activity, but we also need to worry about contamination coming along
with the $^{11}$Be beam. The most worrisome contaminant is $^{11}$Li
that can be ionized by surface ionization in the hot cavity where
laser ionization of Be takes place. The primary production yield is
lower by a factor at least 2000, but the main decay
branch of $^{11}$Li is beta-delayed one-neutron emission that leads to
$^{10}$Be and suppression of $^{11}$Li is therefore needed.

The mass difference between $^{11}$Be and $^{11}$Li, $\Delta M =
Q(^{11}Li) = 20.551$ MeV \cite{mas11}, is sufficiently big that the
resolving power of ISOLDE's high resolution separator HRS, which is up
to $M/\Delta M = 5000$, can easily separate the two nuclei
that have $M/\Delta M = 500$. A tail of $^{11}$Li activity may
nevertheless be present at the $^{11}$Be mass position, but we have 
estimated its magnitude by
looking for the corresponding tails of the nuclei $^9$Li and $^8$Li a
factor $M(^{11}Be)/M(^{11}Li)$ below their nominal setting. In both
cases background prevented a direct identification --- there are no
$\gamma$-rays in either decay, so only the (non-unique)
$\beta$-detection is possible --- but a lower limit of the suppression
factor of $10^3$ was extracted in both cases.
A second source was collected with the separator deliberately set
off-mass by two thirds of the $^{11}$Li-$^{11}$Be mass difference (below
the $^{11}$Be position). The intensity of the $^{11}$Be gamma lines was
here reduced by a factor 500, consistent with the results at the lower masses.

A further reduction can be obtained by blocking collection for the
first 150 ms after the radioactivities are produced by the proton
pulse. Most $^{11}$Li, of halflife 8.5 ms, will have decayed by
then. However, at the end of our collection it was discovered that the
onset of the blocking by mistake was delayed by a few ms allowing up
to 15\% of the $^{11}$Li to leak through. Thus up to $2 \cdot 10^4$
$^{10}$Be could have contaminated the sample through decay of collected $^{11}$Li.

A second possible contaminant could be the BeH molecule $^1$H$^{10}$Be
that would be almost on top of the $^{11}$Be mass since $M/\Delta M =
36600$ for this case. It is quite unlikely that this molecule will be
formed and ionized in sufficient quantities to be a problem, but the
possibility should be kept in mind in future experiments. As will be
seen we do not have indications for any of these contaminants in the
present sample.

\subsection{AMS measurement}

The $^{10}$Be accelerator mass spectrometry (AMS) measurements were
performed at the Tandem Laboratory, Uppsala University, in Sweden
\cite{Ber09}. A dedicated setup based on a NEC 5 MV tandem pelletron
accelerator devoted to high precision and low background detection of
$^{10}$Be was employed. The measurements were carried out according to
the principle presented by Middletone and Klein \cite{Mid86,Mid93}
where ion currents were normalized by $^{17}$O from the $^9$Be$^{17}$O
molecule. The absolute transmission in the AMS-system was determined
with the NIST SRM 4325 standard ($^{10}$Be/$^9$Be= $3.06 \cdot
10^{-11}$ \cite{Mid86}; NIST certificates the ratio to be R$_{SRM} = 2.68 \cdot
10^{-11}$).

The Be-foil used in the investigation was divided into three $10 \times
10$ mm$^2$ samples where one subfoil A was just chemically prepared
and used as a blank value representative for the whole AMS
procedure. Subfoil B and C were both mounted in two aluminium frames
with a circular aperture (diameter 8 mm). Both were installed in the
ISOLDE facility. There was no implantation in subfoil B while
subfoil C was used as the catcher for $^{11}$Be.

All three subfoils were chemically prepared by dissolution in HCl
followed by precipitation to Be(OH)$_2$ by adding NH$_3$ which in turn
was converted to BeO. Finally the BeO was mixed with Nb as a binder
before the AMS-analysis in the accelerator. 

\begin{table}
 \caption{Results of AMS measurements. \label{tab:AMS}}
 \begin{tabular}{cccc}
 \br
 Subfoil & Mass (mg) & R=R$_{foil}$/R$_{SRM}$ & $^{10}$Be ($10^6$ atoms/mg)  \\ \mr
 A (blank) & 3.04 & 0.00149(12) & 2.67(22) \\
 B (background) & 3.65 & 0.00152(11) & 2.73(20) \\
 C (catcher) & 4.22 & 0.00169(13) & 3.03(23) \\
 \br
\end{tabular}
\end{table}

A summary of the results is given in table \ref{tab:AMS}.
No significant difference is observed for the blank and background
foils which indicates no additional $^{10}$Be contamination in the
handling in ISOLDE and during transportation. It is, however,
noteworthy to compare the background values for the foil used with
what is obtained from our normal Be carrier which for 1 mg sample
gives a R $= 0.0005(1)$ corresponding to a factor two lower $^{10}$Be content.

If we use subfoil B as a background to subfoil C the $^{10}$Be signal
can be estimated to be $(1.28 \pm 1.29) \cdot10^6$ atoms. With a total
number of $^{11}$Be nuclei in the sample of $(5.2 \pm 0.3) \cdot
10^{11}$ this leads to a deduced branching ratio of $(2.5 \pm 2.5)
\cdot 10^{-6}$ where the uncertainty is on the one sigma level. There
is therefore no positive evidence for the $^{11}$Be $\beta$p decay,
only an upper limit of the order of the most optimistic theoretical
branching ratios.

\section{The $^8$B experiment}
The $^8$B experiment was carried out at the IGISOL (Ion Guide Isotope
Separator On-Line) separator at the Accelerator Laboratory of
University of Jyv\"{a}skyl\"{a}. The main aim of this experiment was
to determine precisely the neutrino spectrum from the decay of $^8$B
through analysis of the beta-delayed alpha spectrum. The result of
this part of the experiment along with a detailed description of the
experimental procedures is described in \cite{Kir10,Kir11}; the set-up
consisted of four DSSSD detectors each backed by a thick Si detector
in close geometry around the collection point. As a byproduct
the high statistics of the experiment allowed to identify for the
first time an electron capture branch in the decay of $^8$B, namely to
the $2^+$ 16.922 MeV state.  We are here concerned with the possible
electron capture branch to the even higher-lying 17.640 MeV state.

\begin{figure}[ht]
 \hspace{2cm}
 \resizebox{0.7\textwidth}{!}{\includegraphics{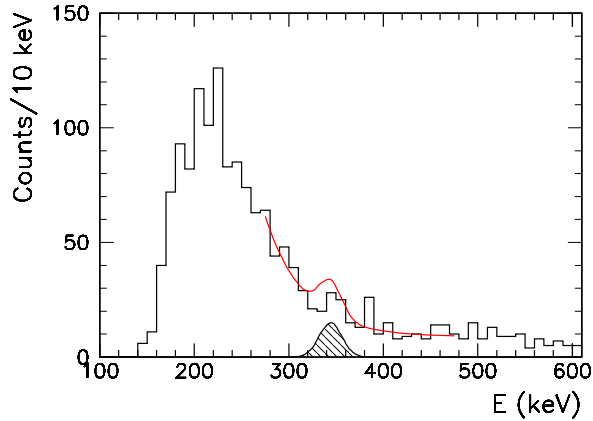}}
\caption{The low-energy spectrum from the decay of $^8$B after
   background suppression. The red curve shows the two sigma 
   upper limit from of a fit  where the peak position was allowed
   to vary within uncertainties.
\label{fig:sp8b}}
\end{figure}

Since the $^8$B experiment recorded a total of 16 million decays, the
expected number of proton detections is 0.36 and we can only report an
upper limit.  The 17.640 keV state will decay by emitting a 337 keV
proton and a 48 keV recoiling $^7$Li ion that could not be detected. The
main background from the decay will be $\alpha$-$\alpha$ coincidences
at low energy or positrons that happen to deposit around 340 keV in
the detector. The latter background can be suppressed by requiring
anti-coincidence with the backing detectors. The former background
needs to be addressed indirectly since the set-up did not include
particle identification. One can suppress the $\alpha$ particles
by doing anti-coincidence with the opposite DSSSD, but to do this
efficiently one needs to restrict the solid angle and only consider
events close to the center of each DSSSD since beta-recoil can cause
the $\alpha$ coincidences to deviate from a linear geometry. This
reduces the effective solid angle by a factor about 9. A small
background nevertheless remains, see figure \ref{fig:sp8b},
and the 95\% confidence level upper limit of the number of counts in a
337 keV peak turns out to be 18. As seen in the figure there are
indications for a peak at 349 keV; however, the uncertainty on the
energy scale is only 7 keV since there were accurately known
calibration lines from the decay of $^{23}$Al close by \cite{Kir11a}.
The 95\% confidence level upper limit on the number of counts in
the 349 keV peak is 45. Combing all factors gives an
upper limit of $2.6 \cdot 10^{-5}$ on the $\beta$p branching ratio at
95\% confidence level. To reach the range of predicted intensity the
experiment needs to improve by a factor 1000. One may gain a factor of
ten by including particle identification, which will reduce the
background in the interesting energy range. Another factor of ten may
be obtainable by increasing the effective solid angle, but one still would
need a factor of ten higher overall production of $^8$B ions.

\section{Outlook}
The theoretical estimates in section 2 showed that beta-delayed proton
emission in the decays of $^{11}$Be and $^8$B is likely to take place
with a branching ratio that can be expected to be a few times
$10^{-8}$. The energies of the emitted proton and the recoiling
nucleus will be small. These two factors make detection of the
processes challenging.

We have carried out first experiments to look for these decay branches
and have obtained upper limits on the branching ratios. These do not
yet reach the level of the theoretical predictions, improvements are
needed by factors of at least $10^2$ and $10^3$ for the two cases.
The continuing improvements of radioactive beam facilities may give
increases in the yields that will allow experiments to reach the
needed level of sensitivity, but other possible improvements should
also be considered. As explained above for $^8$B changes to a more
dedicated set-up should be considered. For the case of $^{11}$Be
the AMS measurement can be considerably improved if the amount of
stable Be in the sample is reduced.  This can be achieved by using a different
catcher foil (e.g.\ Cu, Au or Nb) and after adding a small amount of Be
carrier ($< 0.1$ mg with low $^{10}$Be content) do a refined chemical
separation \cite{Ber09a}. An enhanced sensitivity of two orders of magnitude can
be expected from such improved sample handling. 

Once the decay modes have been established and the branching ratio
securely determined, the next important step will be to determine the
energy distribution of the emitted protons. This will give a more
sensitive test of the theoretical calculations and may in particular
show whether our hypothesis of a direct relation between the $\beta$p
decay mode and the halo structure in these two nuclei is correct.

\bigskip

\ack
We would like to thank I. Mukha and L. Grigorenko for
discussions and F. Sarazin for communicating non-published results on
the $^{11}$Li decay.
This work was partly financed by the Spanish Research funding agency under
project CICYT FPA2007-62170 and FPA2009-07387 and by MINECO through
projects FPA2010-17142 and CPAN CSD-2007-00042.

\end{document}